\documentstyle[12pt]{article}
\begin{document}

\def\beq{\begin{equation}}
\def\eeq{\end{equation}}
\def\bce{\begin{center}}
\def\ece{\end{center}}
\def\bea{\begin{eqnarray}}
\def\eea{\end{eqnarray}}
\def\ben{\begin{enumerate}}
\def\een{\end{enumerate}}
\def\ul{\underline}
\def\ni{\noindent}
\def\nn{\nonumber}
\def\bs{\bigskip}
\def\ms{\medskip}
\def\wt{\widetilde}
\def\brr{\begin{array}}
\def\err{\end{array}}

\hfill UB-ECM-PF-94/5

\hfill March 1994

\vspace*{3mm}

\begin{center}

{\LARGE \bf
One-loop renormalization and asymptotic behaviour of a
higher-derivative scalar theory in curved spacetime}

\vspace{4mm}

\renewcommand
\baselinestretch{0.8}
\medskip

{\sc E. Elizalde}
\footnote{E-mail: eli@ebubecm1.bitnet, eli@zeta.ecm.ub.es} \\
Department E.C.M. and I.F.A.E., Faculty of Physics,
University of  Barcelona, \\ Diagonal 647, 08028 Barcelona, \\
and Center for Advanced Studies, C.S.I.C., Cam\'{\i} de Santa
B\`arbara,
\\
17300 Blanes, Catalonia, Spain \\
{\sc A.G. Jacksenaev} \\
Tomsk Pedagogical Institute, 634041 Tomsk, Russia \\
{\sc S.D. Odintsov}
\footnote{E-mail: odintsov@ebubecm1.bitnet}
\\
Department E.C.M., Faculty of Physics,
University of  Barcelona, \\  Diagonal 647, 08028 Barcelona,
Catalonia,
Spain \\
and Tomsk Pedagogical Institute, 634041 Tomsk, Russia \\
and \
{\sc I.L. Shapiro}
\footnote{E-mail: shapiro@fusion.sci.hiroshima-u.ac.jp}
\\
Tomsk Pedagogical Institute, 634041 Tomsk, Russia \\
and Department of Physics, Hiroshima University, Higashi-Hiroshima
724,
Japan

\renewcommand
\baselinestretch{1.4}

\vspace{5mm}

{\bf Abstract}

\end{center}

A higher-derivative, interacting, scalar field theory in curved
spacetime with the most general action of sigma-model type is
studied.
The one-loop counterterms of the general theory are found. The
renormalization
group equations corresponding to two different, multiplicatively
renormalizable variants of the same are derived. The analysis of
their asymptotic solutions shows that, depending on the sign of one
of the coupling constants, we can construct an asymptotically free
theory which is also asymptotically conformal invariant at strong
(or
small) curvature. The connection that can be established between
one
of the multiplicatively renormalizable variants of the theory and
the
effective theory of the conformal factor, aiming at the description
of quantum gravity at large distances, is investigated.

\vspace{4mm}

\newpage

\ni {\bf 1. Introduction.} \   There are several motivations for
the study of quantum field theories involving higher-derivative
terms. To begin, it is well known in string theory (see \cite{1},
for a review) that if one wants to study the massive higher-spin
modes, one has to modify the standard $\sigma$-model action, by
adding to it an infinite number of terms containing all possible
derivatives. In particular, the terms with two derivatives in
that $\sigma$-model describe the massless modes of the string,
the terms with four derivatives describe the massive higher-spin
modes at first level, and so on.

Moreover, also the effective action of (super)string theory can
be presented under the form of a certain derivative expansion
\cite{1} involving higher-derivative terms of any order. Such
models ---as well as some other higher-derivative gravitational
theories--- often admit singularity-free solutions (for a recent
discussion together with a list of references, see \cite{2,3}),
whose behaviour is, as a rule, in agreement with the violation
of the energy conditions, which is in accordance with the
singularity
theorems \cite{4}.

As a second motivation, higher-derivative theories ---forgetting
about the standard unitarity problem which, most probably, will
be solved only in a non-perturbative approach--- often exhibit
better renormalization properties. A nice example on this line
is quantum $R^2$-gravity (for a review and a list of
references, see \cite{5}), which (unlike Einstein's gravity) is
multiplicatively renormalizable \cite{6} and even asymptotically
free.

Third, a scalar theory with higher-order derivatives, in the
infrared stable fixed point, has been used recently by Antoniadis
and Mottola \cite{7} (see also \cite{8,9}) in order to describe
a truncated version of quantum gravity at large distances. This
theory, which was obtained by integration over the conformal
anomaly \cite{10}, was found by these authors to be
superrenormalizable.

In the present letter we start a systematic study of
higher-derivative scalar theory in curved spacetime.
An action of the
$\sigma$-model type, which is renormalizable in the generalized
sense, will be considered (the scalar field will be chosen to be
dimensionless). The one-loop divergences of the effective action
will be found. For two different variants of this theory (both
of them multiplicatively renormalizable), the corresponding
renormalization (RG) equations to one loop order are
constructed and the asymptotic behaviour of their solutions is
investigated in detail. It is shown that, depending on the sign of
the initial values
 of the coupling constants, the theory under discussion can be
asymptotically free. Moreover, at strong curvature such
asymptotically
free theory becomes also asymptotically conformally invariant.
 \bs

\ni {\bf 2. One-loop divergences.} \   As starting point, let us
impose the condition that the scalar field $\varphi$ be
dimensionless in four-dimensional curved spacetime, namely that
$[\varphi ]=0$. Also, there is just one dimensional constant, which
has dimensions of mass squared, i.e. $[m^2]=2$. Then, dimensional
considerations lead us to the following action (of sigma-model
type and which is renormalizable in the generalized sense)
\bea
S&=& \int d^4x \, \sqrt{-g} \, \left\{ b_1 (\varphi ) \left( \Box
\varphi \right)^2 +  b_2 (\varphi ) \left( \nabla_\mu \varphi
\right) \left( \nabla^\mu \varphi \right)
 \Box \varphi +  b_3 (\varphi ) \left[ \left( \nabla_\mu \varphi
\right) \left( \nabla^\mu \varphi \right) \right]^2 \right. \nn
\\ && +  b_4 (\varphi ) \left( \nabla_\mu \varphi \right) \left(
\nabla^\mu \varphi \right)+  b_5 (\varphi ) +  c_1 (\varphi ) R
\left( \nabla_\mu \varphi \right) \left( \nabla^\mu \varphi
\right) +  c_2 (\varphi ) R^{\mu\nu} \left( \nabla_\mu \varphi
\right) \left( \nabla_\nu \varphi \right) \nn \\
&& + \left.   c_3 (\varphi ) R \Box \varphi + a_1 (\varphi )
R^2_{\mu\nu\alpha\beta }  + a_2 (\varphi ) R^2_{\mu\nu} + a_3
(\varphi ) R^2 + a_4 (\varphi ) R \right\}.
\label{1}
\eea
It is interesting to notice that non-minimal scalar-gravitational
interactions of a new type (with couplings $c_1$, $c_2$ and $c_3$)
appear in action (\ref{1}). Also, in
Eq. (\ref{1}) all generalized coupling constants are
dimensionless, except for $b_4$, $b_5$ and $a_4$, for which we
have:  $[b_4 (\varphi )]=2$,  $[b_5 (\varphi )]=4$,  $[a_4
(\varphi )]=2$. All the possible structures which can appear at
dimension 4 can be obtained from the terms of Eq. (\ref{1}) with
the only help of integration by parts. We must also take
into account
that total derivative terms (such as $\Box^2 a(\varphi )$,
$\Box R$, etc.) have been dropped from Eq.  (\ref{1}) and will
not be considered below. As a further observation, we note that
the action  (\ref{1}) gives a way for obtaining a gravitationally
induced scalar mass and also (if the variant of  (\ref{1})
under discussion is symmetric under $\varphi \rightarrow -
\varphi$) curvature induced spontaneous symmetry breaking at
tree level. This generalizes the corresponding phenomenon which
was found in ref. \cite{12} for the first time in the case of
a theory with a scalar potential of the form
\beq
\lambda \varphi^4 + \xi R \varphi^2.
\eeq

Now we start the discussion of the renormalization structure of
the action (\ref{1}). For the sake of simplicity, in the present
paper we will restrict ourselves to the case where $b_1(\varphi
) \equiv b_1$, $b_2(\varphi ) \equiv b_2$ and $b_3(\varphi ) \equiv
b_3$
are
simply constants (they have no dependence  on $\varphi$). In order
to
obtain the one-loop  effective action divergences,
$\Gamma_{div}$, extremely lengthy and tedious calculations are
necessary. They rely on the use of the background field
method (for a review see \cite{5}) and of Schwinger-De Witt
techniques for the computation of the one-loop effective action
divergences (see \cite{11} for an introduction and \cite{11p} for
the
generalization to the higher derivative case). Details
of this very involved calculation will be given elsewhere. The
result we
have obtained is
\bea
\Gamma_{div} &=&- \frac{2}{\epsilon} \int
 d^4x \, \sqrt{-g} \, \left\{ \frac{5 b_2^2}{4b_1^2} \left( \Box
\varphi \right)^2 + \frac{5 b_2 b_3}{b_1^2} \left( \nabla_\mu
\varphi \right) \left( \nabla^\mu \varphi \right)
 \Box \varphi + \frac{5 b_3^2}{b_1^2} \left[ \left( \nabla_\mu
\varphi \right) \left( \nabla^\mu \varphi \right) \right]^2
\right. \nn \\ && + \left[ \frac{6 b_3 b_4(\varphi ) - 3 b_2
b_4'(\varphi )}{2 b_1^2} - \frac{b_4''(\varphi )}{2 b_1} \right]
\left( \nabla_\mu \varphi \right) \left( \nabla^\mu \varphi
\right) +  \frac{b_4^2(\varphi ) - b_1 b_5''(\varphi )}{2 b_1^2}
\nn \\ && +\left[ \frac{4b_3 \left( -9c_3'(\varphi )+ 9  c_1
(\varphi )+2 c_2 (\varphi )\right)- 8 b_1b_3 - b_2c_2'(\varphi
)}{12 b_1^2} -  \frac{c_1''(\varphi )}{2 b_1} \right] R
\left( \nabla_\mu \varphi \right) \left( \nabla^\mu \varphi
\right)  \nn \\ && +\left[ \frac{2b_3 c_2(\varphi )-b_2^2+b_2
c_2' (\varphi )+4 b_1b_3}{6b_1^2} -  \frac{c_2''(\varphi )}{2 b_1}
\right] R^{\mu\nu}
\left( \nabla_\mu \varphi \right) \left( \nabla_\nu \varphi
\right) \nn \\ && +\left[ \frac{b_2 \left( -9c_3'(\varphi )+ 9
c_1 (\varphi )+2 c_2 (\varphi )\right)- 2 b_1b_2}{6b_1^2} -
\frac{c_3''(\varphi )}{2 b_1} \right] R \Box \varphi \nn \\ &&
- \frac{a_1''(\varphi )}{2 b_1} R^2_{\mu\nu\alpha\beta} + \left[
\frac{c_2^2(\varphi )}{24b_1^2} +\frac{c_2(\varphi )}{6b_1}  +
\frac{1}{30} - \frac{a_2''(\varphi )}{2 b_1} \right] R^2_{\mu\nu}
\nn \\ &&  +\left[ \frac{3 \left( 4c_3'(\varphi )- 4  c_1
(\varphi )- c_2 (\varphi )\right)^2-c_2^2(\varphi )}{96 b_1^2}
+ \frac{2c_3'(\varphi )- 2 c_1 (\varphi )- c_2 (\varphi )}{12
b_1} +\frac{1}{60} \right.
\nn \\ && \hspace{5mm}
- \left.  \frac{a_3''(\varphi )}{2 b_1} \right] R^2
+\left. \left[ \frac{b_4 \left(- 4c_3'(\varphi )+ 4  c_1
(\varphi )+ c_2 (\varphi )\right)}{4b_1^2} - \frac{b_4(\varphi
)}{6 b_1} -  \frac{a_4''(\varphi )}{2 b_1} \right] R \right\}.
\label{2}
\eea
Here $\epsilon = (4\pi)^2 (n-4)$, $\varphi$ is the background
scalar field, and no surface terms have been kept.
Notice that, as one can see, the terms
describing non-minimal interaction of gravity with the scalar field
(i.e.,
those involving $c_1$, $c_2$ and $c_3$) are induced even for
constant
couplings.
To check our results, we have considered the case of a flat
background and there we have made use of standard Feynman graph
analysis (which is also quite involved, due to the presence of
derivatives at the vertices).

Having now the explicit form of the one-loop divergences of the
theory under discussion, we are able to see the structure of its
one-loop renormalization. For doing this we shall consider some
specific variants of action (\ref{1}). The standard conditions
for the one-loop renormalizability of the theory are obtained
from the constraint that the functional form of the one-loop
counterterms  must repeat the form of the corresponding
generalized couplings. These conditions look here as follows (for
$b_1$, $b_2$ and $b_3$ they are trivially satisfied, owing to
the fact that these couplings are constants)
\bea
d_1b_4 (\varphi ) &=& \frac{6b_3b_4(\varphi )-3b_2b_4' (\varphi
)}{2b_1^2} - \frac{b_4''(\varphi )}{2 b_1}, \nn \\
d_2b_5 (\varphi ) &=& \frac{b_4^2(\varphi )-b_1b_5''(\varphi
)}{2b_1}, \ldots
\label{3}
\eea
where $d_1$ and $d_2$ are arbitrary constants, and similarly for
the other couplings. One must solve Eqs. (\ref{3}) for all the
coupling constants in order to obtain the whole class of one-loop
renormalizable theories. However, in this paper we will focus
our interest in  multiplicatively renormalizable theories
only, which form a subclass of the whole class of one-loop
renormalizable scalar theories (\ref{1}).

The analysis of Eqs. (\ref{3}) and the structure of the
divergence index of the theory, as well as dimensional
considerations, show that two variants of the theory given by the
action (\ref{1}) are multiplictively renormalizable, namely the
following ones.
\ms

\ni (1) The theory where all the generalized coupling constants,
except $c_3$, are constants, that is
\bea
&& b_4(\varphi ) = b_4, \ \  b_5(\varphi ) = b_5, \ \
c_1(\varphi ) = c_1, \ \  c_2(\varphi ) = c_2, \ \ c_3(\varphi
) = c_{31}\varphi + c_{32}, \nn \\
&& a_1(\varphi ) = a_1, \ \  a_2(\varphi ) = a_2, \ \
a_3(\varphi ) = a_3, \ \  a_4(\varphi ) = a_4.
\label{4}
\eea
This particular choice of the constants permits us to have the
following conformally invariant 4th order action, which is particular
case of conformal version of the theory (1) \cite{12a}
\beq
\int d^4 x \sqrt{- g}
\varphi \left[ \Box^2 + 2 R^{\mu\nu} \nabla_\mu \nabla_\nu -
\frac{2}{3} R \Box + \frac{1}{3} \left( \nabla^\mu R
\right)\nabla_\mu  \right] \varphi,
\eeq
which corresponds to the following choice of constants
\beq
b_1 = 1, \ \ \ \ \ c_1 = \frac{2}{3}, \ \ \ \ \ c_2 =-2
\label{7}
\eeq
with all the other couplings set equal to zero. Notice that we
could
also put $c_{31}=0$ in (\ref{4}), and we would get in this way
another variant of the above theory.
\ms
\ni (2) The second multiplicatively renormalizable case of the
theory (\ref{1}) is the one given by the following couplings:
\beq
 b_4(\varphi ) = b_4 \, e^{\alpha \varphi}, \ \ \ b_5(\varphi ) =
b_5 \, e^{2\alpha \varphi}, \ \ \  a_4(\varphi ) = a_4 \, e^{\alpha
\varphi},
\label{8}
\eeq
while the remaining generalized coupling constants are chosen as
before, in Eq. (\ref{4}). Notice that this is the theory which,
on a flat background, has been recently used in refs. \cite{7,9} in
order to describe a truncated theory of quantum gravity at large
distances.
\bs

\ni {\bf 3. Renormalization and RG equations.} \   We will here
obtain the RG equations corresponding to the renormalizable
variant (\ref{4}) of the  theory under discussion. In this
section we set $b_1=1$, since the corresponding term plays
the role of the higher-derivative kinetic term.

We must now consider the combination of (\ref{1}) and (\ref{2}),
\beq
S - \Gamma_{div},
\eeq
for the choice of the coupling constants, as in  (\ref{4}) and
(\ref{8}). Expression (\ref{8}) defines the renormalization of
all the coupling constants and scalar field $\varphi$.

First of all, we obtain the one-loop renormalization of the
scalar field in the following form
\beq
\varphi_0 = Z_\varphi^{1/2} \varphi, \ \ \ \ \ Z=1+
\frac{5b^2_2}{2\epsilon}.
\label{9}
\eeq
Using expression (\ref{9}) in order to find the one-loop
renormalization of all the coupling constants (\ref{4}), we are
able to
construct the system of one-loop RG equations for the theory
with couplings given by (\ref{4}):
\bea
\frac{db_2(t)}{dt} = -10b_2(t)b_3(t) + \frac{15}{4} b_2^3(t), &&
b_2(0)=b_2, \nn \\
\frac{db_3(t)}{dt} = -10b_3^2(t) +5 b_2^2(t)b_3(t), &&
b_3(0)=b_3, \nn \\
\frac{db_4(t)}{dt} = -6b_3(t)b_4(t) + \frac{5}{2} b_2^2(t)b_4(t),
&& b_4(0)=b_4, \nn \\
\frac{db_5(t)}{dt} = -b_4^2(t), &&
b_5(0)=b_5.
\label{10}
\eea
This system (\ref{10}) has to be solved in the first place, since
the corresponding running couplings do not change when we come to
flat
space. The RG equations which give the behaviour of the
scalar-gravitational couplings are the following
\bea
\frac{dc_1(t)}{dt} = \frac{2}{3} b_3(t) \left[9c_{31}(t)-9c_1(t)
-2c_2(t) +2 \right]+ \frac{5}{2} b_2^2(t)c_1(t), && c_1(0)=c_1,
\nn \\
\frac{dc_2(t)}{dt} =-\frac{1}{3} \left[ 2c_2(t)b_3(t)-b_2^2(t)
+4b_3(t) \right]+ \frac{5}{2} b_2^2(t)c_2(t), && c_2(0)=c_2, \nn
\\
\frac{dc_{31}(t)}{dt} = \frac{5}{2} b_2^2(t)c_{31}(t), &&
c_{31}(0)=c_{31}, \nn \\
\frac{dc_{32}(t)}{dt} = \frac{1}{3}b_2(t)\left[
9c_{31}(t)-9c_1(t) -2c_2(t) +2 \right]+ \frac{5}{4}
b_2^2(t)c_{32}(t), && c_{32}(0)=c_{32}.
\label{11}
\eea
Finally, the RG equations for the effective vacuum couplings have
the form
\bea
&& \frac{da_1(t)}{dt} =0, \ \ a_1(0) = a_1, \ \ \ \ \ \ \ \
\frac{da_2(t)}{dt} =-\frac{1}{12}c_2^2(t) - 2 c_2(t) -
\frac{1}{15}, \ \ a_2(0) = a_2, \nn \\
&& \frac{da_3(t)}{dt} =\frac{-3[4c_1(t)+c_2(t)-4c_{31}(t)]^2 +
c_2^2(t)}{48} - \frac{2c_{31}(t) -2c_1(t)-c_2(t)}{6} -
\frac{1}{30}, \ \ a_3(0) = a_3, \nn \\
&& \frac{da_4(t)}{dt} =\frac{1}{2}b_4(t) \left[4c_{31}(t)
-4c_1(t)-c_2(t) + \frac{2}{3} \right], \ \ a_4(0) = a_4.
\label{12}
\eea
When comparing with the ordinary versions of the RG equations,
notice that in Eqs. (\ref{10})-(\ref{12}) the following change of
RG
parameter has to be made (with respect to the standard one, $t$):
\
$t \longrightarrow (4\pi)^{-2} t$.
Notice also that, as is usual
in curved space, we adopt the curved spacetime version of the RG
equations (for an introduction, see \cite{5}). In particular, the
UF limit $t \rightarrow \infty$ in flat space, corresponds in
curved spacetime to the strong curvature ---or small distances---
limit (instead of the momentum rescaling $p \rightarrow e^t \, p$,
in curved space the rescaling $g_{\mu\nu} \rightarrow e^{-2t}\,
g_{\mu\nu}$ is used). Notice also that we are not taking into
account
classical dimensions in the RG equations for the dimensional
couplings.
\bs

\ni {\bf 4. Asymptotic solutions.} \ It is remarkable that the RG
equations
above can be solved in an exact way. Its asymptotic analysis
can be carried out completely and we will obtain all possible
asymptotic behaviours of the solutions compatible with the
equations. At the same time, well behaved, approximate solutions
will be found, starting from some initial conditions $b_i(0)$,
$c_i(0)$ and $a_i(0)$.

Eqs. (\ref{10})-(\ref{12}) can be solved
iteratively. In fact, we can start from the two first Eqs.
(\ref{10}),
proceed then with the third (after having substituted the solutions
of
the first two), then with the fourth, and so on.
In particular, these two first equations constitute an exact
differential equation once the integrating factor $\mu = b_2^9 b_3^8$ is
introduced. The equation for the orbit turns out to be a kind of
higer-order hyperbola $12 b_2^{10} b_3^{10} - 5 b_2^{12} b_3^9 =c$,
which is, however, unstable (and, therefore, not relevant for the
asymptotic analysis that follows). All the
possible asymptotic behaviours of the solutions turns out to be rather
simple to find. In fact, as $t \rightarrow +\infty$ only the three
exclusive cases: (i) $b_2^2(t) << b_3(t)$, (ii) $b_2^2(t) >>
b_3(t)$ and (iii) $b_2^2(t) \simeq k \, b_3(t)$, with $k$ finite and
$k \neq 0$, can occur. It is easy to see that the second case is
incompatible (when substituted back into the differential
equations). For the first one,  under the only
conditions that $|b_2|, |b_3| <1$ and that $b_3 >0$ (in order to
prevent singularities from occuring) and after some work, we obtain
\beq
b_2 (t) \simeq \frac{b_2}{1 + 10 b_3 t}, \ \ \ \ \
b_3 (t) \simeq \frac{b_3}{1 + 10 b_3 t}. \label{b23}
\eeq
It is clear that, for any $t >0$, Eqs. (\ref{b23}) are
self-consistent
approximate solutions of the first two Eqs. (\ref{9}), since always
$|b_2(t)|, |b_3(t)| <1$.
Moreover, they are in fact asymptotic solutions, because
$|b_2(t)|, |b_3(t)| \rightarrow 0$, as $t \rightarrow +\infty$.
Thus, as we see, in the ultraviolet limit ($t \rightarrow + \infty$)
the theory is asymptotically free, and this property is obtained by
just choosing the initial value $b_3 >0$. For $b_3 <
0$ we would have the usual zero-charge problem in the UF limit.

In the same way, we obtain
\beq
b_4 (t) \simeq \frac{b_4}{(1 + 10 b_3 t)^{3/5}}, \ \ \ \ \
b_5 (t) \simeq b_5 + \frac{b_4^2}{2b_3} \left[\frac{1}{(1 + 10 b_3
t)^{1/5}} -1\right], \label{b45}
\eeq
with the only additional hypothesis that $|b_4| <1$ ({\it no}
restriction on $b_5$ needs to be imposed).
That is, (\ref{b23}) and (\ref{b45}) are the asymptotic solutions
($t
\rightarrow +\infty$) of Eqs. (\ref{9}), provided
$|b_2|, |b_4| <1$ and
$0<b_3<1$. Moreover, they are also approximate solutions for any
$t>0$.
Notice that we have found an asymptotical solution of a specific
form,
which is self-consistent (when substituted back into the full set
of
Eqs. (\ref{10}))  and asymptotically free.

As for the case (iii) of the above analysis, we may just consider
it asymptotically or either we can impose it rigorously (special
solution of the RG equations)
\beq
b_3(t) = k b_2^2(t), \ \ \ \ k \neq 0,
\label{marg1}
\eeq
and then obtain an  exact solution of the whole system. A detailed
study yields, with the constrains $k=1/4$ and $b_3=b_2^2/4$ for the
initial values, the solution:
\bea
b_2(t) = b_2 \left( 1- \frac{5}{2} b_2^2t \right)^{-1/2}, &&
b_3(t) = \frac{b_2^2}{4} \left( 1- \frac{5}{2} b_2^2t \right)^{-1},
\nn \\ b_4(t) = b_2 \left( 1- \frac{5}{2} b_2^2t \right)^{-2/5}, &&
 b_5(t) =b_5- \frac{2b_4^2}{b_2^2} \left[ \left( 1 \frac{5}{2}
b_2^2t \right)^{1/5}-1 \right].
\label{marg2}
\eea
We again have an asymptotically free type solution, but in the
IR limit only.
In the case that (\ref{marg1}) is only satisfied asymptotically,
then Eqs. (\ref{marg2}) just reflect the behaviour of the $b_i(t)$
for large $t$, with the difference that instead of $b_2$ we must
put a constant, $c$, and that now the above constraint on the
initial values $b_2$ and $b_3$ need not be satisfied.

The analysis of Eqs. (\ref{10}) and (\ref{11}) is carried out in
the same way. For the case (i), the result reads as follows:
\bea
c_1(t) & \simeq & \frac{2(c_2 +2)}{(1 + 10b_3t)^{1/15}} -
\alpha_1,
\ \ \ \ \ \ \alpha_1 \equiv -c_{31} \exp \left( \frac{b_2^2}{4b_3}
\right) - \frac{2}{3}, \nn \\
c_2(t) & \simeq & \frac{c_2 + 2}{(1 + 10b_3t)^{1/15}} -2, \nn \\
c_{31}(t) & \simeq & c_{31}
\exp \left[\frac{b_2^2}{4b_3} \left( 1 -
\frac{1}{1 + 10b_3t} \right) \right], \nn \\
c_{32}(t) & \simeq & c_{32} +\frac{b_2(c_2 +2)}{b_3} \left[
\frac{1}{(1 + 10b_3t)^{1/15}} -1 \right],
\label{ces}
\eea
and
\bea
a_1(t) & = & a_1, \ \ \ \ \ \ \ \ \ \
a_2(t) \  \simeq \ a_2 + \frac{18}{5} t, \nn \\
a_3(t) & \simeq & a_3 - \alpha_2 t, \  \
\alpha_2 \equiv
\frac{1}{4} \left[ 2 c_{31}
\exp \left(\frac{b_2^2}{4b_3} \right) +2\alpha_1 +1\right]^2 +
\frac{1}{6} \left[  c_{31}
\exp \left(\frac{b_2^2}{4b_3} \right) +\alpha_1 +1\right] -
\frac{1}{20}, \nn \\
a_4(t) & \simeq & a_4 + \alpha_3\left[1- \left(1 +
10b_3t\right)^{2/5}\right], \ \ \ \alpha_3 \equiv
\frac{b_4}{2b_3} \left[  - c_{31}
\exp \left(\frac{b_2^2}{4b_3} \right)  - \alpha_1 - \frac{2}{3}
\right].
\label{aes}
\eea
Eqs. (\ref{b23})--(\ref{aes}) are the asymptotic solutions of
Eqs. (\ref{9})--(\ref{11}), for $t \rightarrow + \infty$. The only
requeriments for self-consistency are the ones stablished before:
$|b_2|, |b_4| <1$ and
$0<b_3<1$. Furthermore, they are approximate solutions for {\it
any}
$t>0$. Note also that in the ultraviolet UF limit, the non-minimal
coupling constants
(\ref{ces})  tend to their asymptotically conformal invariant
values (\ref{7}),
independently of their initial values.
Notice that this phenomenon (asymptotical conformal invariance)
has been already found to occur in several asymptotically free GUTs
\cite{5}, but it is somewhat surprising to observe that it can
persist
in higher-derivative theories. It means, for example, that particle
creation of higher-derivative scalar bosons (coming from our model)
in
a Friedmann-Robertson-Walker universe is asymptotically suppressed
at
strong curvature, as it happens for scalar particles in the
asymptotically conformal invariant GUTs \cite{5}. It would be
interesting to understand if this phenomenon holds for the
generalized theory (\ref{1}) as well (in terms of the generalized
RG).
It is also interesting to observe
that the induced running gravitational constant $a_4(t)$ has
similar
behaviour as in the case of GUTs in curved spacetime \cite{5}.

The analysis of the infrared limit
is pretty much the same as that for the
ultraviolet. In fact, the only change in the asymptotic behaviour
concerns the sign of $b_3$, which should be now reversed in order
to avoid the singularity (zero-charge problem). Thus, the final
expressions are exactly
the same as (\ref{b23})--(\ref{ces}) being now the requeriments for
self-consistency: $|b_2|, |b_4| <1$ and
$-1<b_3<0$. Again, with this proviso expressions
(\ref{b23})--(\ref{ces}) are approximate solutions for any $t<0$
big enough in absolute value, and with $-1<b_3 <0$ the theory is
asymptotically free in the IR region (weak curvature limit).
Moreover, the non-minimal scalar graviton couplings, $c_1(t)$,
$c_2(t)$
and $c_3(t)$, tend to their conformal invariant values (\ref{7}) at
weak curvature. Thus, our theory is asymptotically conformal
invariant at weak curvature.

It is intesting to note that asymptotical conformal invariance
may also occur in general theory (1). The values corresponding to the
conformal version of (1) are \cite{12a}
$
b_1(\varphi) = f(\varphi),b_2(\varphi) = f'(\varphi) ,
b_3(\varphi) = b_4(\varphi) = b_5(\varphi) = c_3(\varphi) = a_4(\varphi)
= 0,c_1 = {2 \over3}f(\varphi),c_2 = - 2f(\varphi)
,a_1 = q(\varphi), a_2 = - 2q(\varphi), a_3 = {1 \over3}q(\varphi)$
where $f(\varphi)$ and $q(\varphi)$ are arbitrary functions. It
is natural to suppose that this relations will hold for the one-loop
divergences.
\bs

\ni {\bf 5. RG equations for the conformal sector of quantum
gravity.}
\  Let us now consider the multiplicatively renormalizable theory
which is obtained by choosing the generalized couplings in the form
(\ref{7}). In flat space ($g_{\mu\nu} = \eta_{\mu\nu}$), if we make
the following identification
\beq
b_1 = - \frac{\theta^2}{(4 \pi)^2}, \ \ \ \ b_2 = - 2\zeta \alpha,
\ \ \ \  b_3 = - \zeta \alpha^2,  \ \ \ \ b_4 (\varphi) = \gamma
\exp ( 2\alpha \varphi), \ \ \ \  b_5 (\varphi ) = -
\frac{\lambda}{\alpha^2} \exp (4 \alpha \varphi),
\label{18}
\eeq
we find that our theory corresponds to the one which was used in
Ref. \cite{7} in the infrared stable fixed point $\zeta =0$ to
describe the truncated theory of quantum gravity at large
distances. In (\ref{18}) we have used the same notations as in Ref.
\cite{7}. It is interesting to note that the theory (\ref{18})
admits the interesting vacuum (for simplicity we set $\alpha =1$)
$e^{2\varphi} = \gamma/(2\lambda )$, already at the tree level. The
situation becomes more complicated at one-loop order \cite{9}. It
would be worth, perhaps, to develop some kind of gaussian effective
potential approach (see, for example, \cite{13}) in order to try
understand the non-perturbative structure of this theory.

When considering the specific theory which is obtained with the
choice of coupling constants (\ref{8}) ---and having in mind the
conformal anomaly-induced theory of Ref. \cite{7} as a particular
case of our
general theory--- we come to the conclusion that the coupling
constants $b_2$ and $b_3$ may be connected (in particular, for
$\alpha =1$ they are simply proportional). As one can see from
(\ref{2}), its one-loop renormalization may still respect this
connection ---while it will be destroyed already at one-loop level
if we perform the one-loop renormalization of the scalar field, as
in (\ref{8}), for arbitrary $b_1$. That is why we prefer here not
to do the renormalization of the scalar field, and to consider
instead all coupling constants, including $b_1$ ($\sigma$-model
type renormalization). In this case, the one-loop RG equations can
be easily derived from (\ref{2}):
\beq
\dot{b}_1 =- \frac{5b_2^2}{2b_1^2}, \ \dot{b}_2 =-
\frac{10b_2b_3}{b_1^2}, \ \dot{b}_3 = -\frac{10b_3^2}{b_1^2}, \
\dot{b}_4 = \frac{-6b_3b_4+3\alpha b_2b_4}{b_1^2}+ \frac{\alpha^2
b_4}{b_1}, \  \dot{b}_5 =- \frac{b_4^2}{b_1^2}+ \frac{4\alpha^2
b_5}{b_1},
\label{19}
\eeq
and
\beq
\dot{c}_1 = \frac{2b_3(9c_{31}-9c_1-2c_2)}{3b_1^2}+
\frac{4b_3}{3b_1}, \  \dot{c}_2 = \frac{b_2^2-2b_3c_2}{3b_1^2}-
\frac{4 b_3}{3b_1}, \ \dot{c}_3 = \frac{b_2(
9c_{31}-9c_1-c_2)}{3b_1^2}+ \frac{2b_2}{3b_1}.
\label{20}
\eeq
To simplify the notation, we have not written explicitly the
$t$-dependence in Eqs. (\ref{19}) and (\ref{20}). Also, $\alpha$
and $c_{31}$ are constants, and $b_i(0)=b_i$, $c_j(0)=c_j$.

The vacuum RG equations are
\bea
&& \dot{a}_1 = 0, \ \ \ \ \  \dot{a}_2 =- \frac{c_2^2}{12b_1^2}-
\frac{c_2}{3b_1}- \frac{1}{15}, \nn \\ &&  \dot{a}_3 =-
\frac{3(4c_{31}-4c_1-c_2)^2+c_2^2}{48}- \frac{2c_{31}-2c_1-
c_2}{6b_1}-\frac{1}{30}, \nn \\ &&  \dot{a}_4 = \frac{b_4(
4c_{31}-4c_1-c_2)}{2b_1^2}+ \frac{b_4}{3b_1}+ \frac{\alpha^2
a_4}{2b_1}.
\label{21}
\eea

For simplicity, and having in mind possible applications to the
conformal sector of quantum gravity \cite{7}, we will present here
the solutions corresponding to the matter sector only. Now, the
$t$-dependences will be considered again. The first three equations
(\ref{20}) are equivalent to the following:
\beq
b_3(t) = \frac{b_3}{b_2} b_2(t), \ b_2(t) =  \frac{4b_3}{b_2}
b_1(t)+ b_2 - \frac{4b_1b_3}{b_2}, \ \frac{b_1^2(t)
\dot{b}_1(t)}{\left[b_1(t) -b_1 +b_2^2/(4b_3)\right]^2} =
-\frac{40b_3^2}{b_2^2},
\label{22}
\eeq
where the $t$-dependence has been written explicitly (remember
that the constants are the initial values at $t=0$). As before,
even without being able to solve the sistem of differential
equations exactly, the analysis of all posible asymptotic
behaviours of the solutions can be carried out completely. What
distinguishes now the different possibilites is the behaviour of
$b_1(t)$ as $t\rightarrow \infty$. It can  tend (i) to zero, (ii)
to a non-zero constant, $k$, or (iii) to infinity.The case (i)
turns out to be incompatible. The case (ii) is the most
interesting. The constant $k$ is constrained to be
\beq
k=b_1 -\frac{b_2^2}{4b_3}\equiv b_{1\infty},
\eeq
and we obtain
\bea
b_2(t) \simeq \frac{b_{1\infty}^2b_2}{b_{1\infty}^2+10b_3t}, &&
b_3(t) \simeq \frac{b_{1\infty}^2b_3}{b_{1\infty}^2+10b_3t}, \nn \\
b_4(t) \simeq b_4 e^{b_{1\infty}t}, && b_5(t) \simeq - \frac{
b_4^2}{2b_{1\infty}^3} e^{2b_{1\infty}t} + \left( b_5 + \frac{
b_4^2}{2b_{1\infty}^3}\right) e^{4b_{1\infty}t}.
\label{22p}
\eea
As one can see we found asymptotically free type solutions.
Notice that when $b_{1\infty} <0$ (that is, $b_1 < b_2^2 /(4b_3)$)
both $b_4(t)$ and $b_5(t)$ decay exponentially. The last term of
$b_5(t)$ is then subleading, but we
have included it here in order to show how the initial value $b_5$ is
just washed out for large $t$. Finally, in the case (iii) the
following asymptotic behaviour is easily inferred
\beq
b_1(t) = -\frac{40b_3^2}{b_2^2} \, t + \left(  \frac{b_2^2}{2b_3} -
2 b_1 \right) \ln \left( 1 - \frac{160b_3^3}{b_2^2} \, t \right) +
b_1 + {\cal O} (t^{-1}),
\label{23}
\eeq
valid both for $t \rightarrow + \infty$ (with $b_3>0$) and for  $t
\rightarrow - \infty$ (with $b_3<0$), and similar behaviours for
$b_2$ and $b_3$ (as is clear from (\ref{22})).
On the other hand,
for the remaining $b$-functions we have
\beq
 b_4(t) \simeq b_4 \left( 1-\frac{40b_3^2}{b_1b_2^2} \, t
\right)^{(\alpha_2 b_2^2)/(40b_3^2)}, \label{24} \eeq and \bea
&& b_5(t)  \simeq b_5 \left( 1- \frac{40b_3^2}{b_1b_2^2} \, t
\right)^{(-2\alpha_1^2 b_2^2)/(40b_3^2)} \  \ \ \ \ \ \ \ \ \ \
(b_4(t)
\rightarrow 0), \nn \\
&& b_5(t)  \simeq b_5 +\frac{b_4^2}{b_1(2\alpha_2-40b_3^2/b_2^2)}
\left( 1- \frac{40b_3^2}{b_1b_2^2} \, t \right)^{(\alpha_2
b_2^2)/(20b_3^2) -1} \ \ \ \ \ \  (b_4(t) \rightarrow \infty),
\label{25} \eea
where
\beq
\alpha_1 \equiv \frac{b_2^2}{4b_3} -b_1, \ \ \  \alpha_2 \equiv
\left( \frac{6b_3}{b_2} -3\alpha_1\right) \frac{4b_3}{b_2} -
\alpha_1^2. \eeq
The different asymptotic behaviours for $b_5(t)$
come from the two possible (in principle) behaviours of $b_4(t)$
(depending on the sign of the exponent in (\ref{24})). As one can
see, in this case (iii) the coupling constants do not display an
asymptotically free behaviour.
Notice that using the same prescription for the renormalization of
the scalar theory as in Sect. 3, we would have obtained the same
behaviour for the effective couplings $b_2(t)$ and $b_3(t)$ as in
Sect. 4, while the behaviours of  $b_4(t)$ and $b_5(t)$  would be
quite different.
\bs

\ni {\bf 6. Conclusions.} \   We have discussed in this work the
renormalization of a higher derivative interacting scalar field
theory in curved spacetime. Some variants of this theory have been
shown to be multiplicatively renormalizable and asymptotically
free. The RG structure of the effective theory of the conformal
factor has been also investigated.

There are many interesting directions in which the theory that we
have here introduced deserves further study. One possibility is to
find the one-loop counterterms for the general model (\ref{1})
(with    $b_1(t)$, $b_2(t)$ and $b_3(t)$ being arbitrary functions
of the scalar field) and to construct the generalized RG equations
and investigate their flows. Such approach may be of interest  for
some cosmological applications also, if one would hope to describe
our early universe using a kind of higher-derivative Brans-Dicke
theory (for a recent discussion and a list of references on
scalar-tensor gravity theories, see \cite{14}).
In this case, to complete the study it would be necessary,
as a final step, to discuss the renormalization structure of the
theory (\ref{1}) with the scalar and the gravitational fields being
both quantized. This is a quite complicated program on which we
expect to report elsewhere.

\vspace{5mm}

\noindent{\large \bf Acknowledgments}

SDO would like to thank I. Antoniadis, D. Espriu and R. Tarrach for
helpful discussions, and the members of the Dept. ECM, Barcelona
University, for their kind hospitality.
This work has been  supported by DGICYT (Spain), project No.
PB90-0022, and by CIRIT (Generalitat de Catalunya).

\end{document}